\documentclass[prl,twocolumn,showpacs]{revtex4-1}

\usepackage {amssymb}
\usepackage{graphicx}
\usepackage {color}

\begin{document}

\title []{Bifurcation trees of Stark-Wannier ladders for accelerated BECs in an optical lattice}

\author {Andrea SACCHETTI}

\affiliation {Department of Physics, Informatics and Mathematics, University of Modena e Reggio Emilia
\\Via G. Campi 213/A, Modena - 41125 - Italy}

\email {andrea.sacchetti@unimore.it}

\date {\today}

\begin {abstract} In this paper we show that in the semiclassical regime of periodic potential large enough, 
the Stark-Wannier ladders become a dense energy spectrum because of a cascade of bifurcations  
while increasing the ratio between the effective nonlinearity strength and the tilt of the external field; this 
fact is associated to a transition from regular to quantum chaotic dynamics. \ The sequence 
of bifurcation points is explicitly given.
\end{abstract}

%
%


\maketitle

The dynamics of a quantum particle in a periodic potential under an homogeneous external field is one of 
the most important problems in solid-state physics. \ When the periodic potential is strong enough then we are 
in the semiclassical regime where tunneling between adjacent wells of the periodic potential is practically 
forbidden; in the opposite situation tunneling may occur and the particle performs Bloch oscillations. \ Dynamics 
of particles become more interesting when we take into account the interaction among them, as we must do in 
the case of interacting ultracold atoms. \ In fact, accelerated ultracold atoms moving in an optical lattice 
\cite {Bloch1,Bloch2,RSN,SPSSKP,Shin} has opened the field to multiple applications, as well as the measurements 
of the value of the gravity acceleration $g$ using ultracold Strontium atoms confined in a vertical 
optical lattice \cite {FPST,PWTAPT}, direct measurement of the universal Newton gravitation constant 
$G$ \cite {RSCPT} and of the gravity-field curvature \cite {RCSMPT}. 

Because of the periodicity of the potential associated to the optical lattice, it is expected the existence of 
families of stationary states with associated energies displaced on regular ladders, the so-called 
Stark-Wannier ladders \cite {WS,GKK} (see also \cite {REWWK} for numerical computation of Stark-Wannier states 
for BECs in an accelerated optical lattice); this picture implies, at least for a single particle model, Bloch 
oscillations. \ When one takes into account the binary particle interaction of the condensate nonlinear effects 
occur and new sub-harmonic oscillations appear \cite {S4,KKG,WWMK}. \ More recently, Meinert {\it et al} 
\cite {MMKLWGN} have observed that when the strength of the uniform acceleration is reduced 
a transition from regular to quantum chaotic dynamics is observed; in their experiments evidence of the fact 
that the energy spectrum emerges densely packed, as predicted by \cite {BK} by means of a numerical simulation 
for a lattice with a finite number of wells, is given.

In fact, such a problem has been intensively studied in the recent years by means of numerical methods. \ In 
\cite {LZG} the authors consider a one-dimensional BEC of particles described by the Gross-Pitaevskii equation; 
they reduce the problem to a quasi-integrable dynamical system which displays classical-like Kolmogorov-Arnold-Moser 
structured chaos. \ In \cite {HVZOMB} the authors model the cloud of ultracold bosons in a tilted lattice by 
means of the Bose-Hubbard Hamiltonian that incorporates both the tunneling between neighboring sites and the 
on-site interaction; by means of such an approach they are able to identify regular structures in a globally 
chaotic spectra and the associated eigenstates exhibit strong localization properties in the lattice. \ In 
\cite {KGK} the authors, making use of the mean-field and single band approximations, describe the dynamics of a 
BEC in a tilted optical lattice by means of a discrete nonlinear Schr\"odinger equation; in the strong field limit 
they demonstrate the existence of (almost) non spreading states which remain localized on the lattice region populated 
initially. \ Finally, \cite {VG} can give numerical evidence of the quasi-classical chaos on the emergence of 
nonlinear dynamics.

In this paper we consider the dynamics of ultracold interacting atoms in a periodic potential subjected to 
an external force. \ We can show a transition from the semiclassical picture, where each atom is localized 
on a single well of the periodic potential, to a chaotic picture, for strength of the nonlinearity term large 
enough, associated to a cascade of bifurcations of the energy spectrum; in particular, we can see that when the 
ratio between the effective strength of the nonlinearity interaction term and the strength of the external 
homogeneous field becomes larger of some given values then bifurcations of the stationary solutions occur and 
new stationary solutions localized on a larger number of wells appear. \ In our model the structure of a 
bifurcation trees arising from the Wannier-Stark ladders clearly emerges and the sequence 
of bifurcation points is explicitly given.

Transversely confined BECs in a periodic optical lattice under the effect of the gravitational force are 
governed by the one-dimensional time-dependent Gross-Pitaevskii (GP) equation with a periodic potential and a 
Stark potential
\begin{eqnarray}
i \hbar \partial_t \psi = - \frac {\hbar^2}{2m} \partial_{xx}^2 + V (x) \psi + m g x \psi + \gamma |\psi |^2 \psi \label {Eq1}
\end{eqnarray}
where the BEC's wavefunction $\psi (x,t)$ has constant norm: $\| \psi (\cdot ,t) \|_{L^2} = \| \psi_0 
(\cdot )\|_{L^2}$, where $\psi_0 (x)$ is the initial wavefunction of the BEC, $m$ is the mass of the atoms, 
$g$ is the gravity acceleration, $\gamma$ is the one-dimensional nonlinearity strength and $V(x)$ is the periodic 
potential associated to the optical lattice potential. \ In typical experiments \cite {Bloch1} the periodic 
potential has the usual shape $V(x) = V_0 \sin^2 (k_L x) $ where $b= {\pi}/{k_L}$ is the period and $V_0 = 
\Lambda_0 E_R$ where $E_R$ is the photon recoil energy. 

If one looks for stationary solutions
\begin{eqnarray*}
\psi (x,t) =e^{i \lambda t/\hbar} \psi (x)
\end{eqnarray*}
to the time-dependent GP equation (\ref {Eq1}) it turns out that $\lambda$ is real-valued and that $\psi (x)$ 
is a solution to the time-independent GP equation; then, we may assume that $\psi (x)$ is a real-valued function by 
means of a gauge argument (see Lemma 3.7 by \cite {P}). \ Hence, the time-independent GP equation becomes 
\begin{eqnarray}
\lambda \psi = - \frac {\hbar^2}{2m} \partial_{xx}^2 \psi + V (x) \psi + mg x \psi + \gamma \psi^3  \label {Eq3}
\end{eqnarray}
where $\psi (x)$ is a real valued function. \ First of all let us remark that the stationary solutions to 
eq. (\ref {Eq3}), if there, must be displaced on regular ladders. \ Indeed eq. (\ref {Eq3}) is invariant by 
translation $x \to x+b$ and $\lambda \to \lambda - mg b$, because $V(x+b)=V(x)$ where $b$ is the lattice's 
period. \ Thus we have families of stationary solutions $(\lambda_j , \psi_j (x))$, $j\in {\mathbb Z}$, 
where $\lambda_j = \lambda_0 + j mg b$ and $\psi_j (x) = \psi_0 (x-j b )$ for some $\lambda_0$ and 
$\psi_0(x)$. \ Therefore, we can restrict our analysis to just one \emph {rung} of the ladder and then we 
replicate the obtained results to all other \emph {rungs}.

By means of the tight-binding approach we reduce equation (\ref {Eq3}) to a discrete nonlinear Schr\"odinger 
equation. \ The idea is basically simple \cite {FS} and it consists in assuming that the wavefunction $\psi (x)$, 
when restricted to the first band of the periodic Schr\"odinger operator, may be written as a superposition of 
vectors $u_\ell (x)$ localized on the $\ell$-th well of the periodic potential; i.e. $\psi (x) = \sum_{\ell \in 
{\mathbb Z} } c_\ell u_\ell (x)$, for some $c_\ell $. \ If $u_\ell (x)$ are real-valued functions then the 
parameters $c_\ell$ are real valued too. \ For instance $u_\ell (x) = W_1 (x-x_\ell )$ 
where $W_1 (x)$ is the Wannier function associated to the first band and $x_\ell = \ell b$ is the center of 
the $\ell$-th well. \ Let $\mathbf {c} = \{ c_\ell \}_{\ell \in {\mathbb Z}} \in \ell^2 ({\mathbb Z} )$ be the 
representation of the wave-vector $\psi (x)$ in the tight binding approximation. \ Therefore, 
the tight-binding approach lead us to a system of discrete nonlinear Schr\"odinger equations which dominant terms 
are given by
\begin{eqnarray}
\lambda c_\ell &=& (\lambda_D + mg C_0) c_\ell - \beta (c_{\ell +1} + c_{\ell -1} ) +\nonumber \\ 
&& \ \ + \gamma \| u_0 \|_{L^4}^4 c_\ell^3 + 
mg b \ell c_\ell \, , \ \ell \in {\mathbb Z} \, , \label {Eq3Bis}
\end{eqnarray}
where $\lambda_D$ is the ground state of a single well potential and where $\beta$ is the hopping matrix element 
between neighboring wells, and $C_0 = \int_{{\mathbb R}} x | u_0 (x)|^2 dx$. \ By means of a simple recasting 
$\mu =\lambda - \mu^\star$, $\mu^\star = (\lambda_D + mg C_0 + 2 \beta )$, 
$\nu = \gamma \| u_0 \|_{L^4}^4$ and $f=mg b$, then equation (\ref {Eq3Bis}) takes the form
\begin{eqnarray}
\mu c_\ell = - \beta (c_{\ell +1} + c_{\ell -1} + 2 c_\ell) +\nu c_\ell^3 + 
f\ell c_\ell \, , \ \ell \in {\mathbb Z} \, , \label {Eq4}
\end{eqnarray}
where $c_\ell$ are real-valued and such that $\sum_{\ell \in {\mathbb Z}} c_\ell^2 =1$; the parameter $\nu$ will 
play the role of the effective strength of the nonlinearity interacting term. \ The theoretical question about the 
validity of the nearest-neighbor model (\ref {Eq4}) has been largely debated. \ In particular, numerical 
experiments \cite {AKKS,EHLZCMA} suggest that the nearest-neighbor model properly works when $\Lambda_0 $ is large 
enough, typically $\Lambda_0 \ge 10$. 

Localized modes of the discrete nonlinear Schr\"odinger equation (\ref {Eq4}) have been already studied by \cite {FS, PS, PSM} when 
the external homogeneous external field is absent (i.e. when $f=0$). \ In particular we should mention the contribution 
given by \cite {ABK} where all the solutions obtained in the anticontinuous limit can be classified and where 
bifurcations are observed. \ As far as we know the same analysis is still missing for equation (\ref {Eq4}) when $f \not= 0$. \ We 
look for solutions to the stationary equation (\ref {Eq4}) when $\Lambda_0$ is large enough; in such a case, 
by means of semiclassical arguments, it turns out that $\beta$ becomes small and the stationary solutions are 
close to the ones obtained in the anticontinuum limit of $\beta \to 0$, where (\ref {Eq4}) reduces to 
\begin{eqnarray}
\mu c_\ell = \nu c_\ell^3 + f\ell c_\ell \, , \ \ell \in {\mathbb Z} \, . \label {Eq5}
\end{eqnarray}

When the nonlinear term is absent, that is $\nu =0$, then we simply obtain a family of solutions $\mu_j =f  j$, 
for any $ j \in {\mathbb Z}$, with associated stationary solutions $\mathbf {c} = \pm \{ \delta_{ j}^\ell 
\}_{\ell \in {\mathbb Z}}$. \ In this case we recover the Wannier-Stark ladders \cite {WS,GKK}. 

Assume now that the nonlinear term is not zero, that is $\nu >0$ for argument's sake. \ In general (\ref {Eq5}) 
has finite mode solutions ${\mathbf {c}}^S = \{ c_\ell^S \}_{\ell \in {\mathbb Z}}$, associated to sets 
$S \subset {\mathbb Z}$ (hereafter called solution-sets) with finite cardinality ${\mathcal N} = \sharp S < \infty$, given by 
\begin{eqnarray}
c^S_\ell = 
\left \{
\begin {array}{ll}
 0 & \ \mbox { if } \ell \notin S \\ 
\pm \left [ \frac {\mu^S - f \ell}{\nu } \right ]^{1/2 } & \ \mbox { if } \ell \in S
\end {array}
\right.  \, ,  \label {Eq6}
\end{eqnarray}
with the condition 
\begin{eqnarray}
\frac {\mu^S}{f} > \max S \, , \label {Eq7}
\end{eqnarray}
because we have assumed that $c_\ell^S $ are real-valued and $\nu >0$. \ Furthermore, since the stationary 
problem (\ref {Eq5}) is translation invariant $\ell \to \ell +1$ and $\mu \to  \mu - f$ then we can always 
restrict ourselves to the \emph {rung} of the ladder such that $ \min S=0$, that is the solution-set has the form 
$S= \{ 0, \ell_1 , \, \ldots \, , \ell_{{\mathcal N}-1} \}$ with $0<\ell_1 < \ell_2 < \ldots < 
\ell_{{\mathcal N}-1}$ positive and integer numbers. \ The normalization condition reads
\begin{eqnarray}
1 =  \sum_{\ell \in S} (c_\ell^S)^{2 } = \sum_{\ell \in S} \left [ \frac {\mu^S - f \ell}
{\nu} \right ] \, , \label {Eq8}
\end{eqnarray}
from which it follows that the energy $\mu$ is given by
\begin{eqnarray*}
\mu^S = \frac {\nu}{\mathcal N} + \frac {f}{\mathcal N} \sum_{\ell \in S} \ell  \, . 
\end{eqnarray*}
Hence, condition (\ref {Eq7}) implies the following condition on the solution-set $S$
\begin{eqnarray}
\frac {\nu}{f} > {\mathcal N} \max S - \sum_{\ell \in S} \ell = 
\sum_{\ell \in S} \left [  \max S - \ell \right ]\, . 
 \label {Eq8Bis}
\end{eqnarray}
In order to characterize the solution-sets $S$ let us introduce the complementary set $S^\star$ of $S$ 
defined as follows
\begin{eqnarray*}
S^\star = \{ \ell^\star := \max S - \ell \ : \ \ell \in S \} \, ; 
\end{eqnarray*}
hence condition (\ref {Eq8Bis}) becomes
\begin{eqnarray}
\frac {\nu}{f} > \sum_{\ell^\star \in S^\star} \ell^\star \, . \label {Eq8Quater}
\end{eqnarray}
Let us now denote by ${\mathcal S}^\star (\nu /f)$ the collection of sets $S^\star$ satisfying (\ref {Eq8Quater}); 
let us also denote by ${\mathcal Q}^\star (n)$ the collection of sets of all non negative integer numbers, 
including the number $0$, which sum is equal to $n$, without regard to order with the 
constraint that all integers in a given partition are distinct; e.g. ${\mathcal Q}^\star (1) = \left \{ 
\{ 0,1\} \right \}$, ${\mathcal Q}^\star (2) = \left \{ \{ 0,2\} \right \}$ and ${\mathcal Q}^\star (3) = 
\left \{ \{ 0,3\} ,\, \{ 0,1,2\} \right \}$. \ Hence, by construction 
\begin{eqnarray*}
{\mathcal S}^\star (n+1) =  {\mathcal S}^\star (n) \cup {\mathcal Q}^\star (n) \, . 
\end{eqnarray*}

In conclusion, we have shown that the counting function ${F}(\nu /f)$ given by the number of solution-sets $S$ of 
integer numbers satisfying the conditions (\ref {Eq8Bis}) and such that $\min S =0$, is given by 
\begin{eqnarray}
{F}(\nu /f) = \sum_{0< n < \nu /f} Q(n) \, ; \label {eq10}
\end{eqnarray}
where $Q(n)$ (see Abramowitz and Stegun \cite {AS}, p. 825) gives the number of ways of writing the integer $n$ 
as a sum of positive integers without regard to order with the constraint that all integers in a given partition 
are distinct; e.g. ${F}(3.1)=Q(1)+Q(2)+Q(3) = 1+1+2 =4$. 

%
%
%
It turns out that ${F}(\nu /f)$ grows quite fast, indeed the following asymptotic behavior holds 
true \cite {AS}:
\begin{eqnarray*}
 Q(n)\sim \frac {e^{\pi \sqrt {n/3}}}{4\cdot 3^{1/4}n^{3/4}} \ \mbox { as } \ n \to \infty \, .
\end{eqnarray*} 
Hence
\begin{eqnarray*}
{F}(n) \sim 
\frac {\exp \left [ {\pi} (n/3)^{1/2} \right ] }{2 {\pi} (n/3)^{1/4}} 
\end{eqnarray*}
as $n$ goes to infinity.

\begin{center}
\begin{figure}
\includegraphics[height=7cm,width=8cm]{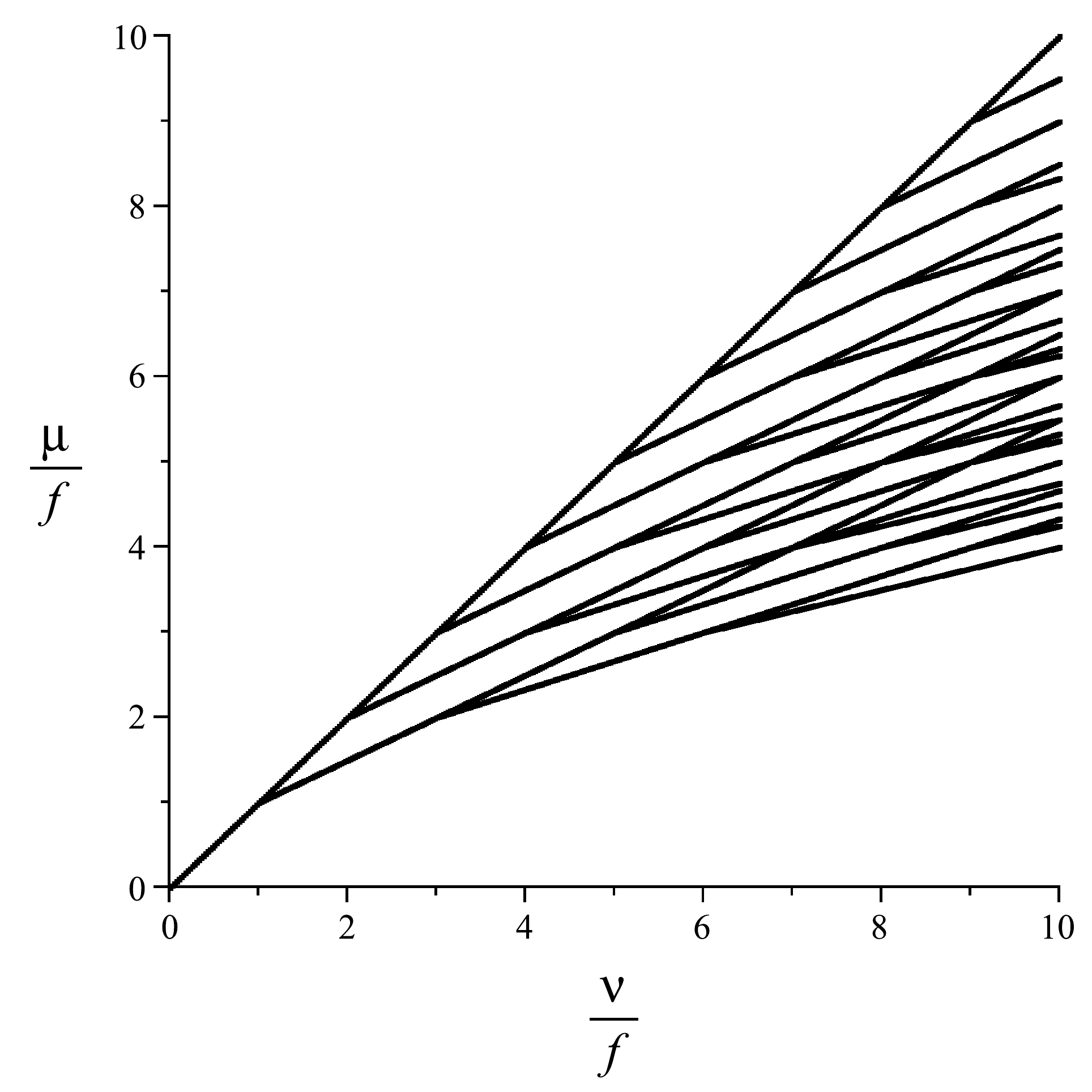}
\caption{\label {Figura2} Here we plot the values of the energy $\mu /f$ associated to stationary solution-sets $S$ 
such that $\min S =0$; we can see a cascade of bifurcations when $\nu/f$ increases. \ This picture occurs for 
each \emph {rung} of the ladder.}
\end{figure}
\end{center}

A cascade of bifurcation points, when $\nu /f$ takes the value of any positive integer, occurs; indeed, when the 
ratio $\nu /f$ becomes larger than a positive integer $n$ then $Q(n)$ new stationary solutions appear. \ This 
fact can be seen in Figure \ref {Figura2}, where we plot the values of the energy ${\mu}$, when $\nu /f$ belongs 
to the interval $[0,10]$, associated to the solution-sets $S$ such that $\min S =0$. \ By translation $\mu 
\to \mu + j f $, $j \in {\mathbb Z}$, we must replicate this picture to the general situation where 
$\min S = j$, $j \in {\mathbb Z}$; that is this picture occurs for each \emph {rung} of the ladder and then 
the collection of values of $\mu$ associated to stationary solutions is going to densely cover the whole real axis.
If one looks with more detail the bifurcation cascade one can see that we have ${\mathcal N}$-mode solutions for 
any value of ${\mathcal N}$. \ For instance, for ${\mathcal N}=1$ we have 1-mode solutions associated to 
solution-sets $S= \{  j \}$, for any $ j \in {\mathbb Z}$, given by $\mu^{\{  j \} } = \nu + f  j$ and 
${\mathbf {c}}^{\{  j\} } = \pm \{ \delta_\ell^{ j} \}_{\ell \in {\mathbb Z}}$. \ That is we recover the 
(perturbed) Wannier-Stark ladder.

For ${\mathcal N}=2$ we have two-mode stationary solutions associated to solution-sets of the form 
$S= \{  j ,  j + \ell_1 \}$ for any $ j \in {\mathbb Z}$ and $\ell_1 \in {\mathbb N}$, where
\begin{eqnarray*}
\mu^{\{  j ,  j + \ell_1 \}} = \frac 12 \nu +  j f + \frac 12 f \ell_1
\end{eqnarray*}
under the condition $\ell_1 > {\nu}/{f}$. \ Therefore, we can conclude that $2$-mode solutions exists only 
if $ {\nu}/{f} >1$, and the elements of the vector ${\mathbf {c}}^{ \{  j ,  j + \ell_1 \} } $ are given by 
\begin{eqnarray*}
c^{ \{  j ,  j + \ell_1 \} }_\ell = 
\left \{ 
\begin {array}{ll} 
 0 & \mbox { if } \ell \not=  j,  j+\ell_1  \\ 
\pm \left [ \frac 12 + \frac 12 \frac {f}{\nu} \ell_1 \right ]^{1/2} & \mbox { if } \ell =  j \\ 
\pm \left [ \frac 12 - \frac 12 \frac {f}{\nu} \ell_1 \right ]^{1/2} & \mbox { if } \ell =  j + \ell_1
\end {array}
\right.
\end{eqnarray*}

In general, ${\mathcal N}$-mode stationary solutions are associated to solution-sets of the form
\begin{eqnarray}
S = \{  j , \  j+\ell_1,\ \ldots \ ,\  j+\ell_{{\mathcal N}-1} \} \label {Eq9}
\end{eqnarray}
where $ j \in {\mathbb Z}$ and $0<\ell_1 < \ell_2< \ldots < \ell_{{\mathcal N}-1} \in {\mathbb N}$, the value 
of $\mu^S$ is given by
\begin{eqnarray*}
\mu^S = \frac {\nu}{{\mathcal N}} +  j f + \frac {f}{{\mathcal N}} \sum_{r =1}^{{\mathcal N}-1} \ell_r 
\end{eqnarray*}
under condition (\ref {Eq7}). \ As a particular family of ${\mathcal N}$-mode solutions we consider  
solution-sets of the form (\ref {Eq9}) for any $ j \in {\mathbb Z}$ and $\ell_{r+1}-\ell_{r} =1$. \ They are 
associated to
\begin{eqnarray*}
\mu^S = \frac {\nu}{{\mathcal N}} + f  j + \frac 12 f ({\mathcal N}-1)
\end{eqnarray*}
and then condition (\ref {Eq7}) implies that 
\begin{eqnarray*}
\frac {{\mathcal N}({\mathcal N}-1)}{2} <  \frac {\nu}{f}  
\end{eqnarray*}
Hence, we can observe a second bifurcation phenomenon: stationary solutions associated to solution-sets 
with ${\mathcal N}$ elements arises from solution-sets with ${\mathcal N}-1$ elements when $\nu /f$ 
becomes bigger than the critical value $(\nu/f)^{\mathcal N} =  {{\mathcal N}({\mathcal N}-1)}/{2} $. 

\begin{center}
\begin{figure}
\includegraphics[height=7cm,width=8cm]{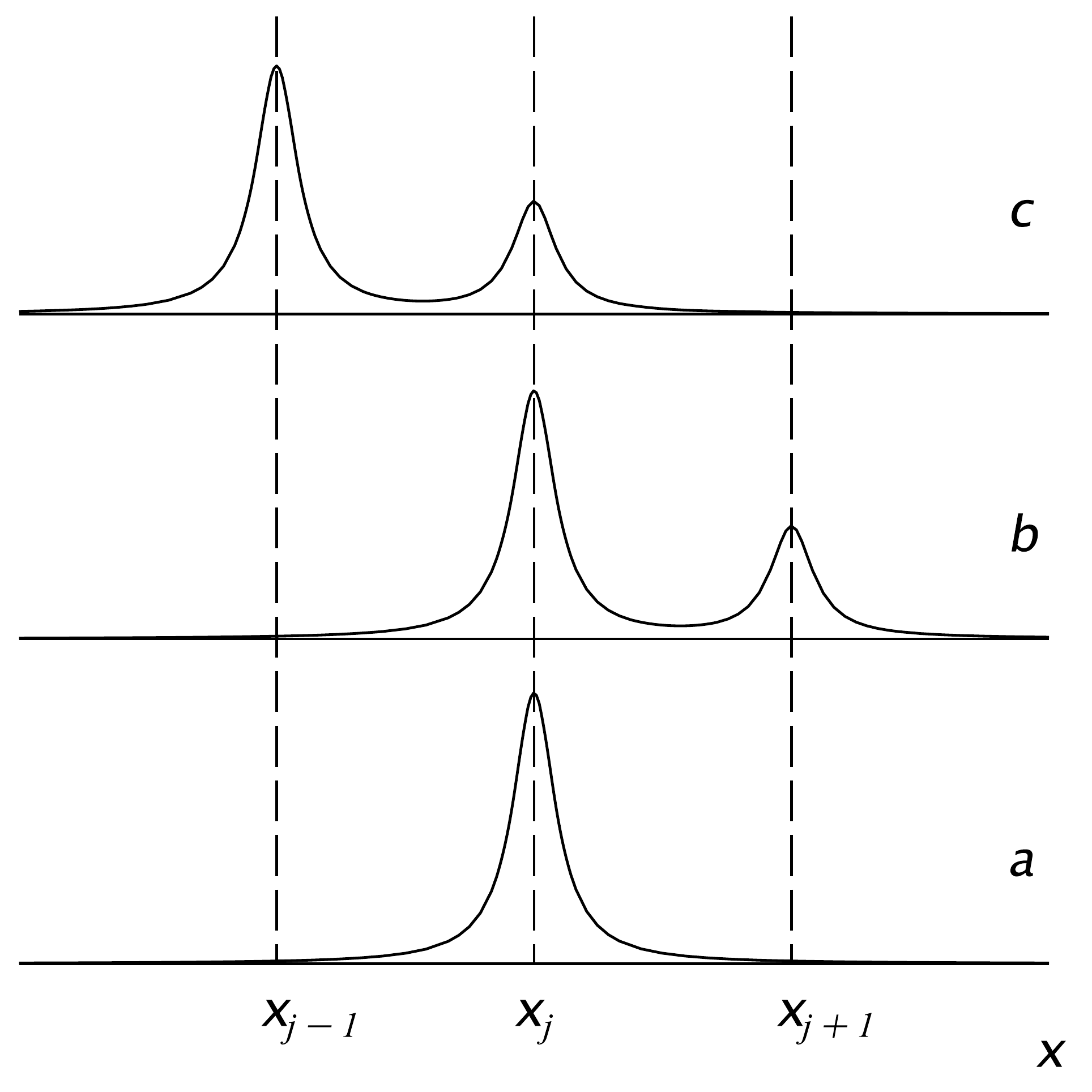}
\caption{\label {Figura3} Here we plot the absolute value of the stationary solutions associated to the 
solution-sets $S_1$ (fig. a), $S_2$ (fig. b) and $S_3$ (fig. c); $x_j$ denotes the center of the $j$-th well of the periodic 
potential.}
\end{figure}
\end{center}
In order to understand the effect of such a stationary solutions on the BEC's dynamics we consider, at first, the 
case where $\nu/f$ is less than one; then we have a family of solutions of the form $ \psi (x,t) = e^{i (\mu + 
\mu^\star ) t/\hbar} u_j (x)$ where $\mu =\nu + j f$ and where $u_j (x)$ is localized on the $j$-th well of the 
periodic potential, $j \in {\mathbb Z}$. \ In fact, in such a case the there is no interaction among these 
solutions, and the density of probability to find the state in the $j$-th well is time independent. \ Let us 
consider now the case when $\nu /f$ is bigger that $1$, i.e. $\nu/f = 3/2$ for argument's sake; then in such 
a case we have that different stationary solutions may be supported on the same well of the periodic potential. \ In 
particular, let us fix our attention on a given well with index $j$, then we have $3$ stationary 
solutions localized on the $j$-th well associated to the solution-sets (see Figure \ref {Figura3})
\begin{eqnarray*}
\begin {array}{ll}
S_1 = \{ j \}; & \mu^{S_1} = \nu + f j  \\ 
S_2 = \{ j,j+1 \}; & \mu^{S_2} = \frac 12 \nu + f j + \frac 12 f  \\ 
S_3 = \{ j-1,j \}; & \mu^{S_3} = \frac 12 \nu +fj -\frac 12 f 
\end {array}
\end{eqnarray*}
If we consider the superposition of these stationary solutions on the $j-$th well then it behaves like
\begin{eqnarray*}
e^{i \mu^\star t /\hbar + i \nu t /2 \hbar + i f j t /\hbar } q (t') u_j (x)
\end{eqnarray*}
where we set $t' = f t/\hbar$ and 
\begin{eqnarray}
q (t') = \left [ c_j^{S_1} e^{i\nu t' /2f } + e^{i t'/2} c_j^{S_2} + e^{-i t'/2} c_j^{S_3} \right ]   \label {Eq12Bis}
\end{eqnarray}
where $c^{S_1}_j = \pm 1$, $c_j^{S_2} = \pm \left [ \frac 56 \right ]^{1/2}$, 
$ c_{j+1}^{S_2} = \pm \left [ \frac 56 \right ]^{1/2}$, $c_{j-1}^{S_3} = 
\pm \left [ \frac 56 \right ]^{1/2}$, $c_j^{S_3} = \pm \left [ \frac 16 \right ]^{1/2}$. \ As a result we observe 
a beating behavior of the density of probability associated to different frequencies; one beating motion has 
period $2\pi$, which is $\nu$-independent and it coincides with the period of the Bloch oscillations, a second 
beating motion has two periods depending on $\nu/f$ given by $T_1 = 4 \pi \left [ 1+ \nu /f \right ]^{-1}$ and 
$T_2 = 4 \pi \left [ -1+ \nu /f \right ]^{-1}$. \ For bigger values of $\nu/f$ then we 
may consider a larger number of stationary solutions which all supports contain a fixed and given well, then the behavior 
on this given well of the 
superposition of such a stationary solutions will be given by means of a periodic function with period $2\pi$, coinciding 
with the Bloch period, plus a large number of periodic functions with different periods; since the number of these 
periodic functions will increase when the ratio $\nu /f$ increases then we expect a chaotic behavior for large $\nu /f$. \ In 
fact, we should underline that a linear combination (like (\ref {Eq12Bis})) of stationary solutions to a nonlinear equation is not, in general, a solution to the same equation. \ However, if 
we consider the limit of small $\nu$ (provided that $\nu/f$ is much bigger than $1$) then we can expect that, for 
fixed times, the contribution due to the non linear perturbation may be estimated and the linear combination of 
stationary solutions approximates a solution to the nonlinear equation.                       

Now, we only have to show that the stationary solution to equation (\ref {Eq5}) obtained in the anticontinuum 
limit goes into a stationary solution to eq. (\ref {Eq4}) when $\beta$ is small enough. \ Indeed, let $\mu^S$ be 
a solution of the anticontinuum limit (\ref {Eq5}), where we can always assume that $\mu^S >0$ by means of the 
translation $\ell \to \ell +1$. \ If we rescale $c_\ell \to \left [ {\mu^S}/{\nu} \right ]^{1/2} c_\ell $ and 
if we set $\beta'= \beta /\mu^S$ and $f'= f/\mu^S$ then the equation (\ref {Eq4}) takes the form 
\begin{eqnarray*}
\left ( 1 - c_\ell^{2} \right ) c_\ell = \beta' ( c_{\ell +1} + c_{\ell -1} + 2 c_\ell )+ f' \ell c_\ell \, . 
\end{eqnarray*}
In conclusion we may extend the solutions to (\ref {Eq5}), obtained in the anticontinuum limit $\beta \to 0$, to 
the solutions to equation (\ref {Eq4}) for $\beta$ small enough if the tridiagonal matrix 
\begin{eqnarray*}
T (\beta ') = \mbox {tridiag} ( \beta' ,  f' \ell -1 + 3 c_\ell^{2} +2\beta' , \beta' ) \, , 
\end{eqnarray*}
obtained deriving the previous equation by $c_\ell$, is not singular at $\beta' =0$, where $c_\ell$ is the solution 
obtained for $\beta' =0$ (see, e.g., Appendix A by \cite {ABK}). \ In particular, it is not hard to see that 
$T (0) = \mbox {diag} (T_\ell)$ has a diagonal form, where $T_\ell =    {f \ell }/{\mu^S} -1 + 3 c_\ell^{2}$ and where 
$c_\ell$ is given by (\ref {Eq6}). \ Hence, a simple straightforward calculation gives that $ \inf_{\ell \in {\mathbb Z}} |T_\ell | >0$. 

In conclusion, in the present contribution we have shown for the first time in the context of BECs in a 
tilted lattice a relevant phenomenon: the occurrence of a cascade 
of bifurcation points in the energy spectrum on the emergence of the nonlinear dynamics, where the associated stationary 
solutions are localized on few lattice's sites. \ This fact gives a theoretical justification of the chaotic 
behavior for large nonlinearity, and it agrees with previous numerical predictions 
\cite {MMKLWGN,BK,LZG,HVZOMB,KGK,VG}. \ We think that the present contribution, with the new result of the 
existence of bifurcation trees, may give a substantially advance in the understanding of the occurrence of 
quasiclassical chaos for BECs in a tilted lattice.

\begin {acknowledgements}
{This work is partially supported by Gruppo Nazionale per la Fisica Matematica (GNFM-INdAM).}
\end {acknowledgements}

\end {document}